\documentclass[12pt,a4paper]{article}
\begin{document}

\def\br{{\bf r}}
\def\bQ{{\bf Q}}

\title{\bf Polarizability of microemulsion droplets\\
}
\author{M. Richterov\'{a} and V. Lis\'{y}\\\\
Department of Biophysics, P. J. \v{S}af\'{a}rik University,\\
Jesenn\'{a} 5, 041 54 Ko\v{s}ice, Slovakia}
\date{}
\maketitle

\begin{abstract}
Spheroidal fluid droplets immersed in another fluid and thermally
fluctuating in the shape are considered. The polarizability of the
droplet is evaluated up to the second order in the fluctuation
amplitudes and also the previous first-order calculations from the
literature are corrected. The correlation functions of the
polarizability tensor components are found and used to describe
the polarized and depolarized scattering of light, and the Kerr
effect on microemulsions in the limit of small concentration of
the droplets. An alternative simple derivation of the Kerr
constant is also given assuming that the droplet in a weak
electric field becomes a prolate ellipsoid. We consider both the
case when the thickness of the surface layer is neglected and when
the droplet is covered by a layer of nonzero thickness. The result
differs significantly from that used in the literature to describe
the Kerr-effect measurements on droplet microemulsions. Due to the
difference the bending rigidity constant of the layer should be
increased about two times in comparison with the value found in
the original experiments.
\end{abstract}

\section{Introduction}
Microemulsions are formed after the addition of surface-active
molecules into the mixture of two immiscible fluids (oil and
water). The surfactants are spread at the oil-water interface as a
dense monolayer. The properties of the layer determine the phase
behavior and thermodynamical stability of
microemulsions~\cite{mic}. Within the Canham-Helfrich concept of
interfacial elasticity~\cite{canham,helfrich}, the surfactant
monolayer is characterized by the bending and saddle-splay modules
$\kappa$ and $\overline{\kappa}$, respectively, the spontaneous
curvature $C_{s}$, the surface tension coefficient $\alpha$, and
the equilibrium radius of the droplet, $R_{0}$. In addition, the
free energy of the droplet is determined by the pressure
difference $\Delta p$ (pressure inside the droplet minus outside).
In real microemulsions there is some distribution of the droplets
in radii. The polydispersity of this distribution, $\varepsilon$,
can be regarded as a microemulsion parameter instead of $\Delta
p$~\cite{borkovec}. The number of these basic parameters lowers in
the case when the microemulsion is in the state of the so called
two-phase coexistence (with the excess amount of the dispersed
droplet phase)~\cite{borkovec}. The determination of the above
parameters has been attempted by a number of experimental
techniques~\cite{modern}. However, different experimental methods
yield very different values of the parameters even for the same
microemulsion systems. For example, there is some one order
difference in the bending rigidity $\kappa$ determined from the
Kerr-effect measurements~\cite{borkeicke} and the neutron
scattering associated with spin echoes~\cite{huang} (for a more
discussion see Ref.~\cite{lisybru}). We have shown in our recent
paper~\cite{lisybru} that the interpretation of the neutron and
light scattering experiments does not correspond to the reality.
In particular, the thermal droplet fluctuations in the shape are
not appropriately taken into account in the description of these
experiments. In Ref.~\cite{lisybru} we calculated the intermediate
scattering function that is used to describe the scattering from
dilute microemulsion and emulsion solutions. The theory
consistently takes into account the droplet shape fluctuations to
the second order in the fluctuation amplitudes. Comparing the
theory and experimental data from the literature, we have found
the microemulsion parameters to be in a notable disagreement with
the values determined in the original experimental works operating
with the previous theories that do not take (or take not
appropriately) into account the droplet fluctuations. So, the
bending rigidities that we have extracted from the experiments are
significantly lower than the values found in the neutron spin-echo
experiments~\cite{huang,far,farago} but larger than possessed by
the spinning drop measurements~\cite{sottmann}, the Kerr
effect~\cite{borkeicke} or a combination of dynamic light and
neutron scattering~\cite{hel1,hel2}. It would be thus useful to
have adequate theoretical description of the different
experimental probes of microemulsions.\\
In the present work the polarizability of a spheroidal droplet is
evaluated. Having a model for the polarizability tensor
$\alpha_{ik}$ of a microemulsion droplet, such experiments like
the Kerr effect or the polarized and depolarized scattering of
light could be described. They could serve as alternative probes
of the droplet shape fluctuations and thus of the microemulsion
parameters. Similar calculations can be already found in the
literature. In the paper~\cite{linden} the polarizability of an
ellipsoid is evaluated. In that work the shape fluctuations have
not been considered and, as shown below, the surface free energy
found there is not correct. In Ref.~\cite{bork} the fluctuations
of a spherical droplet are considered, however, also those
calculations should be corrected. Moreover, they are carried out
only to the first order in the fluctuation amplitudes that is
insufficient in some cases when the observed quantities are
represented by the products of the diagonal polarizability tensor
components. In the next section a brief phenomenological theory of
the shape fluctuations of droplets is given. In the third section
the polarizability tensor $\alpha_{ik}$ is found to the second
order in the fluctuations. Then it is applied to the description
of the Kerr effect on microemulsion and the scattering of light.
In the subsequent section a simple derivation of the Kerr constant
is given assuming that in a weak electric field the fluid droplet
becomes a prolate ellipsoid with small eccentricity. The
derivation is done both for the case when the thickness of the
surface layer of the droplet is negligible and when the droplet is
covered by a membrane with nonzero thickness. The result differs
significantly from that known in the literature~\cite{linden}. Due
to the difference the bending rigidity constant should be
increased about two times in comparison with the value found in
the original experiments~\cite{linden}. In Conclusion, the
obtained results and possible improvements of the theory are
discussed.
\section{Shape fluctuations of spherical droplets}
Consider a flexible droplet taking a spherical shape in
equilibrium. The fluid of the droplet is assumed to be
incompressible and the equivalent-volume radius of the droplet is
$R_{0}$. The instantaneous shape of the deformed droplet can be
described by the deviation of its radius from $R_{0}$, in
spherical harmonics,
\begin{equation} \label{eq1}
f(\vartheta,\varphi)=R(\vartheta,\varphi)/R_{0}-1,
\end{equation}
where
\begin{equation} \label{eq2}
f(\vartheta,\varphi)=\sum_{l,m}u_{lm}(t)Y_{lm}(\vartheta,\varphi),
\end{equation}
with $m=-l, -l+1, ..., l$, and $0\leq l$~\cite{sparling,lebedev}.
When $l=0$ or 1, the coefficients $u_{lm}$ can be expressed as
quadratic combinations of the rest of expansion coefficients, e.g.
\begin{equation} \label{eq3}
u_{00}=-(4\pi)^{-1/2}\sum_{l>1,m}\mid u_{lm}\mid^{2},
\end{equation}
that is a consequence of the constraint on the
droplet volume (the $l=0$ mode corresponds to the overall
"breathing" of the droplet). Analogously, the $l=1$ mode
corresponds (to the second order in $u$) to the translational
motion of the droplet as a whole. The necessary time correlation
functions are as follows:
\begin{equation} \label{eq4} \langle
u_{l0}(0)u_{l0}(t)\rangle=\frac{k_{B}T}{\alpha_{l}R_{0}^{2}(l+2)(l-1)}\exp(-\Gamma_{l}t),
\end{equation}
\begin{equation} \label{eq5} \alpha_{l}=\alpha-2\kappa
C_{s}/R_{0}+\kappa l(l+1)/R_{0}^{2}.
\end{equation}
Here, $\alpha=\sigma+C_{s}^{2}\kappa/2$ ($\sigma$ is the
microscopic interfacial tension~\cite{borkovec}). The decay rates
$\Gamma_{l}$ can be found in our previous paper~\cite{libruza}
where the shape fluctuations of compressible surface layers have
been studied in detail (it is generally believed that the
surfactant monolayer behaves like an almost incompressible
two-dimensional fluid; for $\Gamma_{l}$ in the limit of
incompressible layers see also Ref.~\cite{gurin}).\\
Finally, the distribution of the droplets in
radii as it follows from the phenomenological theory of the
droplet formation~\cite{borkovec} is
\begin{equation} \label{eq6}
f(R_{0})\propto
\exp[-\frac{1}{2\varepsilon}(1-\frac{R_{0}}{R_{m}})^{2}],
\end{equation}
where $R_{m}$ is the mean radius of the droplets.
The generalized Laplace condition~\cite{borkovec} relates the
polydispersity $\varepsilon$ to the characteristics of the layer,
\begin{equation} \label{eq7}
\varepsilon=\frac{k_{B}T}{8\pi
(2\kappa + \overline{\kappa})}.
\end{equation}
Here, for simplicity, the two-phase coexistence is assumed, when
$\alpha=(2\kappa+\overline{\kappa})R_{m}^{-2}= \kappa
C_{s}/R_{m}$. For small $\varepsilon$ the distribution (6) has a
sharp maximum around $R_{m}$, $\langle R_{0}\rangle\approx R_{m}$,
and $\langle(R_{0}-R_{m})^{2}\rangle\approx \varepsilon
R_{m}^{2}$, neglecting small terms $\sim\exp(-1/2\varepsilon)$.\\
For dense microemulsions the interaction between the droplets
should be taken into account. This is still an open question since
it seems that the droplets do not interact like hard
spheres~\cite{weitz,libru}. Most often dilute solutions of
droplets are studied assuming that the effect of interaction is
negligible. The influence of the entropy of dispersion should be
also included into the consideration. It will change the
polydispersity and the quantity $\alpha_{l}$ from Eq. (5). There
is no agreement in the literature as to the concrete expression
for the entropy. Within the random mixing approximation one should
add to the denominator in Eq. (7) a quantity $2k_{B}TF(\Phi)$,
where for small volume fractions $\Phi$ of the droplets
$F\approx\ln \Phi-1$. The mean quadrate of the amplitude of
fluctuations does not explicitly depend on the function $F$,
\begin{equation} \label{eq8}
\langle
u_{l0}^{2}\rangle=\{(l-1)(l+2)[\frac{\kappa}{k_{B}T}l(l+1)-\frac{1}{8\pi\varepsilon}]\}^{-1}.
\end{equation}
\section{Polarizability of a droplet}
As mentioned in Introduction, the polarizability of a fluctuating
droplet was already evaluated by Borkovec and Eicke~\cite{bork}.
However, that work should be corrected in some points. The authors
calculate the polarizability for a fluid droplet of infinite
dielectric constant $\epsilon$ in vacuum. One finds a number of
errors in these calculations. Then the authors remark that the
dipole field generated by a droplet of infinite $\epsilon$ in
vacuum is the same as the dipole field generated by an ellipsoid.
Based on this observation, they write the result for a droplet
with a finite dielectric constant in a dielectric medium simply
using the known result for a dielectric ellipsoid with small
eccentricities~\cite{landau}. In general, such a reasoning is not
correct. In particular, it is not applicable in our problem of
finding the polarizability tensor $\alpha_{ik}, i,k=x,y,z$, of a
droplet, since $\alpha_{ik}$ should be in general calculated at
least to the second order in the fluctuations. This follows from
the fact that the observed quantities correspond to the products
of the polarizability tensor components. Below the polarizability
is evaluated up to the second order in the fluctuation amplitudes.\\
Consider a spheroidal droplet whose shape is described by Eq. (1).
The dielectric constant of the droplet is $\epsilon_{i}$ and the
outer medium is characterized by the constant $\epsilon_{e}$. To
find the polarizability of the droplet, one has to calculate the
electric field generated by the droplet in an external
electrostatic field $\overrightarrow{E_{0}}$. This means to solve
the Laplace equation for the potential $\Phi$ inside and outside
the droplet, together with the boundary conditions at the
interface between the two medii,
\begin{equation} \label{eq9}
\Phi^{(i)}=\Phi^{(e)}, \qquad \epsilon D_{n}^{(i)}=D_{n}^{(e)},
\qquad \textrm{at} \qquad r=R_{0}(1+f),
\end{equation}
where $D_{n}$ is the normal component of electric induction,
$\epsilon=\epsilon_{i}/\epsilon_{e}$, and the indices $i$ and $e$
refer to the interior and exterior of the droplet. At infinity the
resulting electric intensity becomes $\overrightarrow{E_{0}}$. Let
the initial field is oriented along the axis $z$. We then search
for the solution in the form $$\Phi^{(i)}=-\frac{3}{\epsilon+2}r
\cos\vartheta +r\sum_{M}b_{M}^{(z)}Y_{1M},$$
\begin{equation}
\label{eq10}
\Phi^{(e)}=-r\cos\vartheta+\frac{\epsilon-1}{\epsilon+2}r^{-2}\cos\vartheta+r^{-2}
\sum_{M}a_{M}^{(z)}Y_{1M,}
\end{equation}
where we temporarily reduced the variables by replacing
$r/R_{0}\rightarrow r$ and $\Phi/E_{0}R_{0}\rightarrow \Phi$. That
is, the field is represented by a potential due to a perfect
sphere plus a small addition due to the distortion from the
spherical shape. Such a deformation is described by the terms
containing small coefficients $a_{M}^{(z)}$ and $b_{M}^{(z)}$.
Only the dipole field is considered. To satisfy the second
boundary condition in Eqs. (9) one has first to find the normal
vector to the deformed droplet interface. The normal is defined
through the vectors
$\overrightarrow{r}_{\vartheta}=\partial_{\vartheta}\overrightarrow{r}$
and $\overrightarrow{r}_{\varphi}=
\partial_{\varphi}\overrightarrow{r}$, using Eq. (1),
\begin{equation} \label{eq11}
\overrightarrow{n}=\frac{\overrightarrow{r}_{\vartheta}\times
\overrightarrow{r}_{\varphi}}
{(\overrightarrow{r}_{\vartheta}\times
\overrightarrow{r}_{\varphi})^{2}}.
\end{equation}
Performing the calculation we obtain, to the second order in small
$f$,$$\overrightarrow{n}\overrightarrow{\nabla}\Phi
=(\overrightarrow{\nabla}\Phi)_{r}\{1-\frac{1}{2}[(\frac{\partial
f}{\partial \vartheta})^{2} +\sin^{-2}\vartheta(\frac{\partial
f}{\partial \varphi})^{2}]\}$$
\begin{equation} \label{eq12}
+(\overrightarrow{\nabla}\Phi)_{\vartheta}(f-1)\frac{\partial
f}{\partial \vartheta}+ \sin^{-1}\vartheta
(\overrightarrow{\nabla}\Phi)_{\varphi}(f-1)\frac{\partial
f}{\partial \varphi}.
\end{equation}
When $\Phi^{(i)}$ and $\Phi^{(e)}$ from Eqs. (10) are substituted
in Eq. (12), the second boundary condition from Eq. (9) becomes
\begin{equation} \label{eq13}
\epsilon\sum_{M}b_{M}^{(z)}Y_{1M}=-2\sum_{M}a_{M}^{(z)}Y_{1M}+
3\frac{\epsilon-1}{\epsilon+2}[2f\cos\vartheta
+\frac{\partial f}{\partial \vartheta}\sin\vartheta].
\end{equation}
Together with the condition of continuity of the potential,
\begin{equation} \label{eq14}
\sum_{M}b_{M}^{(z)}Y_{1M}=\sum_{M}a_{M}^{(z)}Y_{1M}-
3\frac{\epsilon-1}{\epsilon+2}f\cos\vartheta,
\end{equation}
one obtains, to the first order in $f$, the following
equation for the determination of the coefficients $a_{M}^{(z)}$:
\begin{equation} \label{eq15}
\sum_{M}a_{M}^{(z)}Y_{1M}=3\frac{\epsilon-1}{\epsilon+2}
[f\cos\vartheta+\frac{\sin\vartheta}{\epsilon+2}
\frac{\partial f}{\partial \vartheta}].
\end{equation}
Multiplying this equation by $Y_{1M}$ and integrating over all
angles $\vartheta$ and $\varphi$, one obtains the desired
coefficients $a_{M}^{(z)}$. This can be easily done expressing the
products $Y_{lm}Y_{1m'}$ that appear in the integrals through sums
of spherical harmonics. These sums always contain the
Clebsch-Gordan coefficients $(l100\mid10)$~\cite{edwards} that are
nonzero only for $l=0$ or $l=2$. The $l=0$ mode is excluded since
it gives corrections of the second order in $u_{lm}$ or becomes
zero when differentiated with respect to $\vartheta$ (the second
term in Eq. (15)). We thus have only the spherical harmonics of
order 1 and 2 so that the integration is performed in elementary
functions. In this way we find from Eqs. (14) and (15)
\begin{equation} \label{eq16}
a_{M}^{(z)}=\frac{3}{\sqrt{5}}(\frac{2}{\sqrt{3}}\delta_{M0}+\delta_{M1}+\delta_{M,-1})
(\frac{\epsilon-1}{\epsilon+2})^{2}u_{2M},
\qquad b_{M}^{(z)}=-\frac{3}{\epsilon-1}a_{M}^{(z)}.
\end{equation}
Quite similarly the response of the droplet can be considered when
the external field is oriented along the axes $x$ and $y$. In Eqs.
(10) one has just to replace $z=r\cos\vartheta$ by
$x=r\sqrt{2\pi/3}(Y_{1,-1}-Y_{11})$ and
$y=ir\sqrt{2\pi/3}(Y_{1,-1}+Y_{11})$, and repeat the calculations.
Instead of the coefficients $a_{M}^{(z)}$ we obtained
\begin{equation} \label{eq17}
a_{0}^{(x)}=\frac{3}{\sqrt{10}}(\frac{\epsilon-1}{\epsilon+2})^{2}(u_{2,-1}-u_{21}),
\qquad
a_{\pm1}^{(x)}=\pm\sqrt{\frac{3}{10}}
(\frac{\epsilon-1}{\epsilon+2})^{2}(u_{20}-\sqrt{6}u_{2,\pm2}),
\end{equation}
\begin{equation} \label{eq18}
a_{0}^{(y)}=-i\frac{3}{\sqrt{10}}(\frac{\epsilon-1}{\epsilon+2})^{2}(u_{2,-1}+u_{21}),
\qquad
a_{\pm1}^{(y)}=-i\sqrt{\frac{3}{10}}
(\frac{\epsilon-1}{\epsilon+2})^{2}(u_{20}+\sqrt{6}u_{2,\pm2}).
\end{equation}
The relation between the coefficients $b_{M}^{(i)}$ and
$a_{M}^{(i)}$ is the same as for $i=z$ in Eq. (16). The set of the
obtained coefficients $a_{M}$ and $b_{M}$ fully determines the
dipolar field of a droplet in the first approximation in the
droplet fluctuations. To find the second-order correction to this
solution, we act in the following way. We represent the searched
coefficients as $a_{M}\rightarrow a_{M}+\Delta_{M}$ and
$b_{M}\rightarrow b_{M}+\delta_{M}$, where $\Delta$ and $\delta$
are of the second order in the amplitudes $u$. Substituting the
solutions (10) in Eqs. (9) using (12), the two boundary conditions
are obtained for the unknown corrections $\Delta$ and $\delta$.
Combining the two equations we obtain
$$\sum_{M}\Delta_{M}^{(z)}Y_{1M}=\frac{2\epsilon-3}{\epsilon-1}
\sum_{M}a_{M}^{(z)}fY_{1M}-\frac{4\epsilon-1}{(\epsilon-1)(\epsilon+2)}
\sum_{M}a_{M}^{(z)} [\frac{\partial Y_{1M}}{\partial\vartheta}
\frac{\partial f}{\partial \vartheta}+\frac{1}{\sin^{2}\vartheta}
\frac{\partial Y_{1M}}{\partial\varphi} \frac{\partial
f}{\partial\varphi}]$$
\begin{equation} \label{eq19}
-3\frac{(\epsilon-1)
(\epsilon+4)}{(\epsilon+2)^{2}}f^{2}\cos\vartheta-6\frac{\epsilon-1}{(\epsilon+2)^{2}}
\sin\vartheta\frac{f\partial f}{\partial\vartheta}
+3\frac{\epsilon-1}{\epsilon+2}u_{00}Y_{00}\cos\vartheta.
\end{equation}
Here, $a_{M}^{(z)}$ are from Eq. (16). There is no need to search
for the full solution of this equation. All experimentally
observed quantities that we construct using the solution for the
potential $\Phi$ have to be in the final step averaged over the
fluctuations $u$. Having this in mind, and since we are interested
in the solution correct to the second order in the fluctuations,
we can perform the averaging already in Eq. (19). By this way we
obtain the solution to Eq. (19) in a simplified form that however
gives correct contributions to the averaged quantities of the
second order in the fluctuations:
\begin{equation} \label{eq20}
\Delta_{\pm1}^{(z)}=0, \qquad \Delta_{0}^{(z)}\equiv
\Delta=-\sqrt{\frac{3}{\pi}}\frac{\epsilon-1}{(\epsilon+2)^{2}}
[3\frac{(\epsilon+1)(\epsilon+11)}{\epsilon+2}u_{20}^{2}
+(\epsilon+3)\sum_{l>2}(2l+1)u_{l0}^{2}].
\end{equation}
Analogously, quadratic corrections can be obtained in the cases
when the external field is oriented along the axes $x$ and $y$.
The change of the corresponding coefficients $a_{M}$ is as
follows:
\begin{equation} \label{eq21}
\Delta_{0}^{(x)}=0, \qquad
\Delta_{\pm1}^{(x)}=\mp\frac{1}{\sqrt{2}}\Delta, \qquad
\Delta_{0}^{(y)}=0, \qquad
\Delta_{\pm1}^{(y)}=\frac{i}{\sqrt{2}}\Delta.
\end{equation}
Now it is easy to obtain the polarizability tensor components,
that is the main purpose of the paper. Writing the solution (10)
for $\Phi^{(e)}$ through the cartesian coordinates $x, y, z$, from
the expression for the dipole field
$\Phi^{(e)}=\overrightarrow{d}\overrightarrow{r}/r^{3}$, the $x,
y, z$ components of the dipole moment are
\begin{equation}
\label{eq22}
\overrightarrow{d}=E_{0}R_{0}^{3}\{\sqrt{\frac{3}{8\pi}}(a_{-1}^{(z)}-a_{1}^{(z)}),
-i\sqrt{\frac{3}{8\pi}}(a_{-1}^{(z)}+a_{1}^{(z)}),
\frac{\epsilon-1}{\epsilon+2}+\sqrt{\frac{3}{4\pi}}a_{0}^{(z)}\},
\end{equation}
where the proper dimension is recovered. Comparing this expression
with the definition of the
polarizability,$$d_{i}=\sum_{k}\alpha_{ik}E_{0k},$$ and using Eqs.
(16) and (21), one obtains the polarizability tensor components
$\alpha_{iz}, i=x, y, z$ in the laboratory frame. Analogously the
rest of the components of the tensor $\alpha_{ik}$ is obtained
with the use of Eqs. (17), (18), and (21). The result is as
follows:$$\alpha_{xy}=\alpha_{yx}=-i\frac{3}{2}
\sqrt{\frac{3}{10\pi}}R_{0}^{3}(\frac{\epsilon-1}{\epsilon+2})^{2}(u_{2,-2}-u_{22}),$$
$$\alpha_{xz}=\alpha_{zx}=\frac{3}{2}\sqrt{\frac{3}{10\pi}}R_{0}^{3}(
\frac{\epsilon-1}{\epsilon+2})^{2}(u_{2,-1}-u_{21}),$$
$$\alpha_{yz}=\alpha_{zy}=-i\frac{3}{2}\sqrt{\frac{3}{10\pi}}R_{0}^{3}
(\frac{\epsilon-1}{\epsilon+2})^{2}(u_{2,-1}+u_{21}),$$
$$\alpha_{xx}=R_{0}^{3}\frac{\epsilon-1}{\epsilon+2}[1+\frac{3}{2}
\sqrt{\frac{3}{10\pi}}\frac{\epsilon-1}{\epsilon+2}
(u_{2,-2}+u_{22}-\sqrt{\frac{2}{3}}u_{20})-\widetilde{\Delta}],$$
$$\alpha_{yy}=R_{0}^{3}\frac{\epsilon-1}{\epsilon+2}
[1-\frac{3}{2}\sqrt{\frac{3}{10\pi}}\frac{\epsilon-1}{\epsilon+2}
(u_{2,-2}+u_{22}+\sqrt{\frac{2}{3}}u_{20})-\widetilde{\Delta}],$$
\begin{equation} \label{eq23}
\alpha_{zz}=R_{0}^{3}\frac{\epsilon-1}{\epsilon+2}
[1+\frac{3}{\sqrt{5\pi}}\frac{\epsilon-1}{\epsilon+2}
u_{20}-\widetilde{\Delta}], \qquad \widetilde{\Delta}\equiv
-\frac{1}{2}\sqrt{\frac{3}{\pi}}\frac{\epsilon+2}{\epsilon-1}\Delta,
\end{equation}
where $\Delta$ is introduced in Eq. (20). In the first
approximation with respect to the account of fluctuation our
expressions correct those from Ref.~\cite{bork} where there were
found for $\epsilon\rightarrow\infty$. The dipole moment induced
by an external field is the same as the dipole moment of an
ellipsoid with the main half-axes
$a=R_{0}[1+(e_{x}^{2}+e_{y}^{2})/6]$,
$b=R_{0}[1+(e_{y}^{2}-2e_{x}^{2})/6]$, and
$c=R_{0}[1+(e_{x}^{2}-2e_{y}^{2})/6]$, where the eccentricities
$e_{x}$ and $e_{y}$ are
$e_{x}^{2}=\sqrt{15/2\pi}(-u_{22}+\sqrt{3/2}u_{20})$, and
$e_{y}^{2}=\sqrt{15/2\pi}(u_{22}+\sqrt{3/2}u_{20})$, in the frame
connected with the droplet and with the axes along the main axes
of the ellipsoid. The depolarization coefficients of such an
ellipsoid, $n^{(i)}=R_{0}^{3}/3\alpha_{ii} -1(\epsilon-1)$,
$i=x,y,z$, are $3n^{(z)}=1-(3/\sqrt{5\pi})u_{20},
 3n^{(y)}=1+(3/2\sqrt{5\pi})(u_{20}+\sqrt{6}u_{22})$, and
$3n^{(x)}=1+(3/2\sqrt{5\pi})(u_{20}-\sqrt{6}u_{22})$, that follows
from the general formula for the dipole moment of an ellipsoid
placed in an external field parallel to the axis
$i$~\cite{landau}. The contributions of the second order of the
fluctuation amplitudes change only the diagonal components of the
polarizability tensor. Thus the polarizability anisotropy, that is
reflected e.g. in the Kerr effect, is determined solely by the
ellipsoidal fluctuations (the $l=2$ modes, as already pointed out
in Ref.~\cite{bork}). The higher order terms are determined by all
kinds of the droplet vibrations with $l>1$. Outside the droplet
the resulting electric field is a sum of the applied field and a
field of an electric dipole in the origin with a dipole moment
(when averaged over the fluctuations) $\langle d
\rangle=d_{sph}[1-\langle\widetilde{\Delta}\rangle]$ parallel to
the applied field. Inside the droplet the mean field is oriented
along $\overrightarrow{E_{0}}$ and its absolute value is larger
than that of a perfect sphere. This follows from the solution (10)
for $\Phi^{(i)}$, that gives $\langle E_{x}\rangle = \langle
E_{y}\rangle = 0, \langle
E_{z}\rangle=3E_{0}/(\epsilon+2)-E_{0}\sqrt{3/4\pi} \langle
b_{0}^{(z)}\rangle$, where $\langle b_{0}^{(z)}\rangle =
\langle\delta_{0}^{(z)}\rangle<0$ (if $b$ is calculated to the
second order in fluctuations) is easily found using Eq. (16) and
the continuity of the potential.
\section{The Kerr effect}
The obtained polarizability of a droplet can be used
for a simplest description of the Kerr effect on droplet
microemulsions. When the droplet is placed in an electric field,
the difference between the refractive indices $n_{\parallel}$
parallel and $n_{\perp}$ perpendicular to the field can be
expressed in terms of the optical polarizabilities as
\begin{equation} \label{eq24}
\Delta n\equiv n_{\parallel}-n_{\perp}\approx
\frac{3}{2R_{0}^{3}}n_{e}\Phi(\alpha_{\parallel}^{opt}-\alpha_{\perp}^{opt}),
\end{equation}
where $\Phi$ is the volume fraction of the droplets and $n_{e}$ is
the refractive index of the microemulsion continuous phase. Eq.
(24) follows from the Lorentz-Lorenz formula simplified for the
case of low $\Phi$~\cite{jackson}. To obtain the statistically
averaged quantity $\langle \Delta n\rangle$, we use the full free
energy of a dielectric body in an electric field~\cite{landau},
\begin{equation} \label{eq25}
F-F_{0}=-\frac{1}{8\pi}\int
\overrightarrow{E_{0}}(\overrightarrow{D}-\epsilon_{e}\overrightarrow{E})dV,
\end{equation}
where $F_{0}$ is the free energy of the field without a dielectric
body, and $\overrightarrow{E}$ is the field changed by the
presence of the body. Equation (25) is especially suitable since
we have to integrate only within the volume of the droplet.
Finding the electric intensity inside the droplet and performing
the integration, one obtains
\begin{equation}
\label{eq26}
F-F_{0}=-\epsilon_{e}\frac{\epsilon-1}{\epsilon+2}\frac{R_{0}^{3}E_{0}^{2}}{2}
[1+\frac{3}{\sqrt{5\pi}}\frac{\epsilon-1}{\epsilon+2}u_{20}].
\end{equation}
Using the expansion $\exp[-(F-F_{0})/k_{B}T]$ to the first order
in $u_{20}$ and the polarizability tensor components from Eq. (23)
(with $\epsilon=n^{2}=(n_{i}/n_{e})^{2}$ for the optical
polarizabilities), we finally find the Kerr constant
\begin{equation} \label{eq27}
K=\frac{\langle\Delta
n\rangle}{E_{o}^{2}\Phi}=\frac{81}{40\pi}\frac{R_{0}^{3}n_{e}\epsilon_{e}}{k_{B}T}
(\frac{\epsilon-1}{\epsilon+2})^{2}(\frac{n^{2}-1}{n^{2}+2})^{2}\langle
u_{20}^{2}\rangle.
\end{equation}
This equation agrees with that obtained in~\cite{bork}. Using the
distribution (6), after the averaging over the droplet radii
$R_{0}^{3}$ has to be replaced by $\langle R_{0}^{3}\rangle
\approx R_{m}^{3}(1+3\varepsilon)$. The estimation of the bending
rigidity value obtained from the experiments~\cite{borkeicke} (see
also Ref.~\cite{erratum}) is $\kappa\approx 1 k_{B}T$. For the
discussion of this result see, however, Conclusion.
\section{Depolarized scattering of light}
The effects of
polarization anisotropy are well revealed in the experiments on
the depolarized scattering of light~\cite{pecora}. Let the
scattered field is propagating in the $x$ direction, and the
initial field has a polarization
$\overrightarrow{n_{i}}=\widehat{z}$. Then the intensity of the
depolarized light ($\overrightarrow{n_{f}}=\widehat{y}$) is
\begin{equation} \label{eq28}
I_{VH}=N\langle
\alpha_{yz}^{opt}(0)\alpha_{yz}^{opt}(t)\rangle F_{s}(Q,t).
\end{equation}
Here, $F_{s}$ is the self-diffusion correlation function
of the droplet, $Q$ is the wave-vector transfer at the scattering,
$N$ is the number of droplets in the scattering volume, and
$\langle\rangle$ denotes the thermal equilibrium average over the
ensemble of droplets in the absence of any field. The polarized
component of the scattered light is
\begin{equation} \label{eq29}
I_{VV}=N\langle \alpha_{zz}^{opt}(0)\alpha_{zz}^{opt}(t)\rangle
F_{s}(Q,t).
\end{equation}
Using Eq. (23), one finds
\begin{equation}
\label{eq30}
I_{VH}=\frac{27N}{20\pi}R_{0}^{6}(\frac{n^{2}-1}{n^{2}+2})^{4}\langle
u_{20}^{2}\rangle \exp(-\Gamma_{2}t)F_{s}(Q,t).
\end{equation}
This expression differs from the equation found in
Ref.~\cite{bork} that contains an extra factor $n_{e}^{4}$.
Analogously the intensity of the polarized scattering can be
found,
\begin{equation} \label{eq31}
I_{VV}=NR_{0}^{6}(\frac{n^{2}-1}{n^{2}+2})^{2}[1+\frac{9}{5\pi}(\frac{n^{2}-1}{n^{2}+2})^{2}
\langle
u_{20}^{2}\rangle\exp(-\Gamma_{2}t)-2\langle\widetilde{\Delta}(n)\rangle]F_{s}(Q,t).
\end{equation}
For the integral intensity of the scattering we have, in agreement
with the formula for cylindrically symmetric
molecules~\cite{pecora}, $I_{VV}=I_{ISO}+\frac{4}{3}I_{VH}$, where
$I_{ISO}=N\alpha^{2}$ is the isotropic part of the scattering
determined by the trace $\alpha$ of the polarizability tensor, and
is easily found from Eq. (23). One thus obtains for the
depolarization ratio
\begin{equation} \label{eq32}
\frac{I_{VH}}{I_{ISO}}=\frac{27}{20\pi}(\frac{n^{2}-1}{n^{2}+2})^{2}\langle
u_{20}^{2}\rangle.
\end{equation}
The account for the second-order terms in fluctuations is
necessary in the determination of the polarized and isotropic
scattering. For example, for the system studied in
Ref.~\cite{borkeicke} (water - AOT - \textit{n}-hexane
microemulsion) with the parameters $n_{e}\approx 1.37, n \approx
1, \sqrt{\varepsilon}\approx 0.12$, and $\kappa\approx 1 k_{B}T$,
the isotropic part of the scattering is determined by
$I_{ISO}=NR_{0}^{6}(n^{2}-1)^{2}(n^{2}+2)^{-2}(1-2\langle
\widetilde{\Delta}\rangle)$ with
$\langle\widetilde{\Delta}\rangle\approx 0.4$. In
$\langle\widetilde{\Delta}\rangle$ itself the account for the
$l>2$ modes is important: it represents about $1/3$ of the $l=2$
contribution. Unfortunately, we have no knowledge about
experiments where the depolarized and polarized light scattering
on microemulsions were measured.
\section{A simple derivation of the Kerr constant} In this
section we give a simple alternative derivation of the Kerr
constant (27). First, consider a fluid droplet assuming that the
thickness of the surface layer of the droplet is negligible if
compared to its radius. When such a droplet of the radius $R_{0}$
is placed in a weak electric field $\overrightarrow{E}_{0}$
directed along the axis $z$, it becomes a prolate ellipsoid with
the half-axes, to the second order of the small eccentricity
$e=\sqrt{1-b^{2}/a^{2}}$,
\begin{equation}
\label{eq33} a=R_{0}(1+e^{2}/3),   \qquad b=c=R_{0}(1-e^{2}/6).
\end{equation}
Within the Helfrich model of interfacial
elasticity~\cite{canham,helfrich} the free energy of such an
ellipsoid (without the electrostatic energy) is~\cite{borkovec}
\begin{equation}
\label{eq34} F=-\Delta pV+\sigma A+\int
dA[\frac{\kappa}{2}(c_{1}+c_{2}-2/R_{s})^{2}+
\overline{\kappa}c_{1}c_{2}].
\end{equation}
Here $V$ is the (constant) volume of the droplet, $\Delta p$ is
the pressure inside minus outside the droplet, and $\sigma$ is the
microscopic surface tension. The integral over the surface $A$ of
the ellipsoid yields the bending energy of the droplet. It is
determined through the local curvatures $c_{1}$, $c_{2}$, and the
spontaneous curvature radius $R_{s}$. Performing the integration
over the ellipsoid, one finds
\begin{equation}
\label{eq35} F=F_{0}+\frac{8\pi}{45}e^{4}R_{0}^{2}(\alpha -
\frac{4\kappa}{R_{0}R_{s}}+\frac{6\kappa}{R_{0}^{2}}),
\end{equation}
where $F_{0}$ is for the sphere, and $\alpha$ is now the
macroscopic surface tension for the plane interface from Eq. (5).
The full free energy is obtained adding the energy of the
ellipsoid in the electric field $E_{0}$. The electrostatic energy
is~\cite{landau}
\begin{equation}
\label{eq36}
F_{el}=-\frac{V\epsilon_{e}}{8\pi}\frac{\epsilon-1}{1+n^{(z)}(\epsilon-1)}E_{0}^{2}.
\end{equation}
The depolarization coefficient is $n^{(z)}\approx
[1-4(a-b)/5R_{0}]/3$, so that we have
\begin{equation}
\label{eq37}
F_{el}\approx-\frac{R_{0}^{3}}{2}\epsilon_{e}\frac{\epsilon-1}{\epsilon+2}
[1+\frac{2}{5}\frac{\epsilon-1} {\epsilon+2}e^{2}]E_{0}^{2}.
\end{equation}
Minimalizing the full free energy with respect to the eccentricity
we find
\begin{equation}
\label{eq38}
e^{2}=\frac{9\epsilon_{e}}{16\pi}(\frac{\epsilon-1}{\epsilon+2})^{2}
(\alpha -
\frac{4\kappa}{R_{0}R_{s}}+\frac{6\kappa}{R_{0}^{2}})^{-1}R_{0}E_{0}^{2}.
\end{equation}
To describe the Kerr birefringence, we now need the optical
polarizabilities perpendicular $\alpha_{\perp}^{opt}
=\alpha_{xx}^{opt}$ and parallel
$\alpha_{\parallel}^{opt}=\alpha_{zz}^{opt}$ to the external
field. They are obtained from the expressions for the dipole
moment of the ellipsoid with small eccentricity~\cite{landau},
\begin{equation} \label{eq39}
\alpha_{\perp}^{opt}\approx \frac{3V}{4\pi}\frac{n^{2}-1}{n^{2}+2}
[1-\frac{e^{2}}{5}\frac{n^{2}-1}{n^{2}+2}],    \qquad
\alpha_{\parallel}^{opt}\approx \frac{3V}{4\pi}\frac{n^{2}-1}
{n^{2}+2}[1+\frac{2e^{2}}{5}\frac{n^{2}-1}{n^{2}+2}],
\end{equation}
with $n$ being the relative refractive index from Eq. (27). The
difference in the refractive indices parallel and perpendicular to
the field is~\cite{jackson},
\begin{equation}
\label{eq40} n_{\parallel}-n_{\perp}\approx
\frac{9}{10}n_{e}(\frac{n^{2}-1}{n^{2}+2})^{2}\Phi e^{2}.
\end{equation}
and the Kerr constant is
\begin{equation}
\label{eq41} K=\frac{n_{\parallel}-n_{\perp}}{E_{0}^{2}\Phi}.
\end{equation}
Substituting here the eccentricity from Eq. (38), we finally
obtain
\begin{equation}
\label{eq42}
K=\frac{81}{160\pi}R_{0}n_{e}\epsilon_{e}(\frac{\epsilon-1}{\epsilon+2})^{2}
(\frac{n^{2}-1}{n^{2}+2})^{2} (\alpha -
\frac{4\kappa}{R_{0}R_{s}}+\frac{6\kappa}{R_{0}^{2}})^{-1},
\end{equation}
that agrees with Eqs. (27) and (4, 5). For the case of two-phase
coexistence in microemulsions~\cite{borkovec} Eq. (42)
significantly simplifies according to the formulas from Section 2.
\section{The Kerr effect on droplets covered with a shell}
The simple method used in the preceding section can be readily
generalized for the description of the Kerr effect on the droplets
covered with a vesicle membrane or a surfactant shell of nonzero
thickness. To do this we need only the expressions that generalize
Eqs. (39) for the polarizabilities of the ellipsoid, taking into
account the size of the surface shell. Such expressions have been
found in the work~\cite{linden}. Following that work we assume the
vesicle fluid core of radius $R_{w}$ to be characterized by the
dielectric constant $\epsilon_{w}$, and the continuous phase of
surrounding fluid by the constant $\epsilon_{o}$. For simplicity
and in order to make a comparison with the experiment, we use
$\epsilon_{w}\gg \epsilon_{o}$ which is true when $\epsilon_{o}$
stays for oil and $\epsilon_{w}$ for water, that corresponds to
the experiments~\cite{linden}. A generalization to other, more
complicated cases, is straightforward; the corresponding formulae
for the polarizabilities can be found in the work~\cite{linden}.
The surface shell can consist of two parts: a polar part which is
characterized by the constant thickness $D_{\epsilon}$, and an
apolar part of the thickness $D-D_{\epsilon}$. The polar part is
described by the dielectric constant $\epsilon_{\beta}$,
characterizing the orthogonal (to the surface) components of the
dielectric constant, and by $\epsilon_{\gamma}$ for the parallel
components. Both $\epsilon_{\beta}$ and $\epsilon_{\gamma}$ are
large compared to $\epsilon_{o}$. The apolar part of the layer has
the dielectric constant approximately the same as for the oil.
Then the parallel component of the polarizability tensor of such
an ellipsoid is as follows~\cite{linden}:
$$\alpha_{\parallel}=
R_{w}^{3}\frac{\epsilon_{w}-\epsilon_{o}}{\epsilon_{w}+2\epsilon_{o}}[1+3ae^{2}
\frac{\epsilon_{w}-\epsilon_{o}}{\epsilon_{w}+2\epsilon_{o}}]
+R_{w}^{2}D_{\epsilon}\frac{3\epsilon_{o}}{(\epsilon_{w}+2\epsilon_{o})^{2}}$$
\begin{equation}
\label{eq43}
\times\{{\epsilon_{w}^{2}(\frac{1}{\epsilon_{o}}-\frac{1}{\epsilon_{\beta}})[1+2ae^{2}
\frac{\epsilon_{w}-7\epsilon_{o}}{\epsilon_{w}+2\epsilon_{o}}]+2(\epsilon_{\gamma}-
\epsilon_{o})[1+2ae^{2}\frac{4\epsilon_{w}-\epsilon_{o}}{\epsilon_{w}+2\epsilon_{o}}]}\},
\end{equation}
where $a=2/15$. For $\alpha_{\perp}$ the same expression is valid
but with $a=-1/15$. In the calculation of the eccentricity $e$ we
use the above mentioned inequalities for the dielectric constants
that gives
\begin{equation}
\label{eq44} \alpha_{\parallel}\approx
R_{w}^{3}(1+\frac{2}{5}e^{2})+3R_{w}^{2}D_{\epsilon}
[1+\frac{4}{15}e^{2}+2\frac{\epsilon_{o}\epsilon_{\gamma}}{\epsilon_{w}^{2}}
(1+\frac{16}{15}e^{2})].
\end{equation}
Using this expression we find the electrostatic part of the free
energy of the ellipsoid, which is now instead of Eq. (36)
\begin{equation}
\label{eq45}
F_{el}=-\frac{1}{2}\alpha_{\parallel}\epsilon_{o}E_{0}^{2}.
\end{equation}
Minimalizing the full free energy $F+F_{el}$, with $F$ from Eq.
(35), the eccentricity is
\begin{equation}
\label{eq46}
e^{2}\approx\frac{9}{16\pi}\epsilon_{o}R_{w}E_{0}^{2}\alpha_{2}^{-1}
(1+2\frac{D_{\epsilon}}{R_{w}}),
\end{equation}
where $\alpha_{2}$ is from Eq. (5) with $R_{0}=R_{s}$. From Eq.
(40), rewriting the polarizabilities $\alpha_{\parallel}$ and
$\alpha_{\perp}$ from Eq. (43) for the optical case simply
changing the static dielectric constants by the squares of the
refractive indices~\cite{linden}, we finally obtain$$
\frac{n_{\parallel}-n_{\perp}}{E_{0}^{2}}\approx
\frac{27}{40}R_{w}^{4}\rho n_{o}\epsilon_{o} (\alpha -
\frac{4\kappa}{R_{w}R_{s}}+\frac{6\kappa}{R_{w}^{2}})^{-1}(1+2\frac{D_{\epsilon}}{R_{w}})
\{(\frac{n_{w}^{2}-n_{0}^{2}}{n_{w}^{2}+2n_{o}^{2}})^{2}$$
\begin{equation}
\label{eq47}
+\frac{2Dn_{o}^{2}}{R_{w}(n_{w}^{2}+2n_{o}^{2})^{3}}[n_{w}^{4}(n_{o}^{-2}-n_{\beta}^{-2})(n_{w}^{2}-7n_{o}^{2})
+2(n_{\gamma}^{2}-n_{o}^{2})(4n_{w}^{2}-n_{o}^{2})]\}.
\end{equation}
The difference between this result and the result by Van der
Linden \textit{et al.}~\cite{linden} is significant. This is
because of the difference in the surface energy of the deformed
droplet in the electric field: they have in the first bracket in
Eq. (47) only the term $6\kappa/R_{w}^{2}$. The rest terms are,
however, not negligible if compared with this one: for a detailed
discussion we can refer, e.g., to the work~\cite{ruckenstein}.
Neglecting the surface energy associated with the surface tension
is justified only for an absolutely free vesicle membrane with
identical fluids inside and outside it, but not in other cases. As
well, in general one cannot assume $R_{w}/R_{s}\ll1$ and drop out
the corresponding terms as it was done in~\cite{linden}. So, for a
microemulsion droplet, the two radii can be comparable, e.g. in
the case of two-phase coexistence we have
$R_{w}/R_{s}\approx (2\kappa+\overline{\kappa})/2\kappa$.\\
In the paper~\cite{linden} a detailed comparison between the
theory and the Kerr effect experiment was done from which the
value $\kappa\approx 0.46 kT$ has been extracted. Taking into
account the above discussed improvement of the theory, it is seen
that this value of the rigidity constant is essentially
underestimated. Really, let us express~\cite{lisybru} in Eq. (47)
\begin{equation}
\frac{\alpha R_{w}^{2}}{6} - \frac{2\kappa
R_{w}}{3R_{s}}+\kappa\approx
(\frac{\kappa}{kT}-\frac{1}{48\pi\varepsilon})kT,
\end{equation}
where $\varepsilon$ is the polydispersity of the droplets in
radii. Exactly this expression should be used in the analysis of
the experimental data~\cite{linden} that yielded the value
$\kappa\approx 0.46 kT$. One can see that the lower the
polydispersity in the sample is, the higher value of $\kappa$
would be determined from the experiment. For example, for a
typical polydispersity index~\cite{erratum}
$\sqrt{\varepsilon}=0.12$ one obtains $\kappa\approx 0.92 kT$: a
value two times larger than that found in Ref.~\cite{linden} for
the water - AOT - isooctane droplet microemulsion and very close
to that determined from the Kerr effect measurements by Borkovec
and Eicke~\cite{borkeicke,erratum}.

\section{Conclusion} In the present work the polarizability of a
droplet has been calculated. It was assumed that the shape of the
droplet fluctuates in time and the result for the polarizability
was obtained to the second order in the amplitudes of the
fluctuations. This could be important when the relevant quantities
are expressed through the correlation functions of the diagonal
components of the polarizability tensor, like in the scattering of
light. Of course, the account for the second order in fluctuations
is unnecessary when the polarizability anisotropy is responsible
for the measured effect. We proceeded from the solution of the
Laplace equation for a fluctuating droplet with a finite
dielectric constant in a dielectric medium. We have corrected the
expressions for the polarizability found in Ref.~\cite{bork} where
it was calculated to the first order in the fluctuations and for a
droplet with infinite dielectric constant in vacuum. The obtained
formulae were applied to the description of the Kerr effect and
the depolarized and polarized scattering of light. The expression
for the specific Kerr constant is the same as in Ref.~\cite{bork},
a significant difference has been found in the expressions for the
intensity of the scattered light. We also gave a simple
alternative derivation of the Kerr constant for the case when the
thickness of the surface layer can be neglected and when it is
nonzero. The latter result corrects that from Ref.~\cite{linden}.
A comparison of the theoretical results with the Kerr-effect
experiment on microemulsions gave an estimation of the bending
rigidity constant of about $1 k_{B}T$ for microemulsions
consisting of droplets with relatively large
radii~\cite{borkeicke,linden}. However, this estimation should be
considered with serious doubts. First, the experimental error in
obtaining the Kerr constant by extrapolation of the data to zero
concentration of the droplets is large so that the estimation is
not very reliable. In Ref.~\cite{borkeicke} the radius of the
droplet was determined by standard dynamic light scattering (DLS)
experiments. It is known that the DLS technique is rather
problematic in the determination of microemulsion characteristics
(see the discussion in Ref.~\cite{lisybru}), especially it
concerns the radius of the droplets. It is always larger than the
radius obtained from other techniques like the scattering of
neutrons. Since the signal measured in the Kerr-effect experiments
is sensitive to the radius, it should be determined with a high
precision. Moreover, the polydispersity of the droplet
distribution in radii becomes very important. In
Ref.~\cite{linden} the polydispersity was not determined at all.
In the work~\cite{borkeicke} it was first assumed for the
polydispersity that $\sqrt{\varepsilon}$ (from small-angle neutron
scattering experiments by other authors) varies from about 0.25 to
0.30. In Erratum to Ref.~\cite{borkeicke} the value for
$\sqrt{\varepsilon}$ was changed to about 0.12, based on reports
from the literature on experiments using light scattering
techniques. To our opinion, all the characteristics should be
determined in one series of experiments on the same system. From
available techniques the small-angle neutron scattering seems to
be the method in which the basic characteristics of the
microemulsion droplets are well fixed. Other experimental
techniques, like the Kerr-effect measurements or light scattering
methods, could serve as alternative probes for these
characteristics. For these purposes, however, a question
concerning the role of the interaction between the droplets in
microemulsion should be solved. We note this problem since
according to the recent investigation~\cite{edwards2} a relevant
theory of the electro-optical measurements on microemulsions
should incorporate many-particle correlations even in the case of
small concentrations of the droplets. It appears that long-range
anisotropic density correlations resulting from dipolar
interactions have to be taken into account in a generalization of
the simple single-body theory presented here.\\

\textbf{Acknowledgment}. This work is a part of the PhD. thesis by
M\'{a}ria Richterov\'{a}. It was supported by the grant VEGA No.
1/7401/00, Slovak Republic.

\end{document}